\documentclass[preprint2]{aastex}

\newcommand{\myemail}{ishida@phys.metro-u.ac.jp}
\newcommand{{\scyg}}{SS~Cyg}
\newcommand{{\ch}}{CH~Cyg}
\newcommand{{\ci}}{CI~Cam}
\newcommand{{\groj}}{GRO~J1655$-$40}
\newcommand{{\gin}}{GS~2023+338}
\newcommand{{\vcyg}}{V404~Cyg}
\newcommand{{\xmm}}{\it XMM-Newton}
\newcommand{{\xt}}{XTE~J0421+560}
\newcommand{{\bh}}{BH}
\newcommand{{\ns}}{NS}
\newcommand{{\whd}}{WD}
\newcommand{{\asca}}{\it ASCA}
\newcommand{{\batse}}{BATSE}
\newcommand{{\ro}}{\it ROSAT}
\newcommand{{\chan}}{\it Chandra}
\newcommand{{\rxte}}{\it RXTE}
\newcommand{{\cgro}}{\it CGRO}
\newcommand{{\sax}}{\it BeppoSAX}
\newcommand{{\exo}}{\it EXOSAT}
\newcommand{{\ka}}{K$\alpha$}
\newcommand{{\mekal}}{\sc mekal}

\shorttitle{Possibility of a White Dwarf in CI Cam}
\shortauthors{Ishida et al.}

\begin{document}


\title{Possibility of a White Dwarf as the Accreting Compact Star \\
    in CI Cam ( = XTE J0421+560)}


\author{Manabu Ishida}
\affil{Department of Physics, Tokyo Metropolitan University,\\
1-1 Minami-Osawa, Hachioji, Tokyo 192-0397, Japan}
\email{\myemail}

\and

\author{Kazuyuki Morio and Yoshihiro Ueda}
\affil{Japan Aerospace Exploration Agency,\\
Institute of Space and Astronautical Science,\\
3-1-1 Yoshinodai, Sagamihara, Kanagawa 229-8510, Japan}
%% \email{morio@astro.isas.jaxa.jp, ueda@astro.isas.jaxa.jp}


%\altaffiltext{1}{Visiting Astronomer, Cerro Tololo Inter-American Observatory.
%CTIO is operated by AURA, Inc.\ under contract to the National Science
%\altaffiltext{2}{Society of Fellows, Harvard University.}
%\altaffiltext{3}{present address: Center for Astrophysics,
%    60 Garden Street, Cambridge, MA 02138}
%\altaffiltext{4}{Visiting Programmer, Space Telescope Science Institute}
%\altaffiltext{5}{Patron, Alonso's Bar and Grill}


\begin{abstract}
We present results from {\asca} observations of the binary {\ci}
both in quiescence and in outburst
in order to identify its central accreting object.
The quiescence spectrum of {\ci} consists of soft and hard components
which are separated clearly at aound 2--3~keV.
A large equivalent width of an iron {\ka} emission line
prefers an optically thin thermal plasma emission model to
a non-thermal power-law model for the hard component,
which favors a white dwarf as the accreting object, 
since the optically thin thermal hard X-ray emission is 
a common characteristic among cataclysmic variables
(binaries including an accreting white dwarf).
However, since the power-law model,
which represents the X-ray spectrum of the soft X-ray transients in quiescence,
provides with an equally good fit to the hard component statistically,
we cannot exclude possibilities of a neutron star or a black hole
from the quiescence data.

The outburst spectrum, on the other hand, is composed of a hard component
represented by a multi-temperature optically thin thermal plasma emission
and of an independent soft X-ray component that appears
below 1~keV intermittently on a decaying light curve of the hard component.
The spectrum of the soft component is represented well by a blackbody 
with the temperature of $0.07-0.12$~keV
overlaid with several K-edges associated with highly ionized oxygen.
This, together with the luminosity as high as
$\sim 1\times 10^{38}{\rm erg\;s}^{-1}$,
is similar to a super-soft source (SSS).
The outburst in the hard X-ray band
followed by the appearance of the soft blackbody component
reminds us of recent observations of novae in outburst.
We thus assume the outburst of {\ci} is that of a nova,
and obtain the distance to {\ci} to be 5--17~kpc by means of the relation
between the optical decay time and the absolute magnitude.
This agrees well with a recent estimate of the distance of 5--9~kpc
in the optical band.
All of these results from the outburst data prefer a white dwarf for the
central object of {\ci}.

\end{abstract}




\section{Introduction}

On 1998 Mar 31 \citet{smi98} discovered the new X-ray transient {\xt}
with the All-Sky Monitor (ASM; \citet{lev96})
on board {\it Rossi X-ray Timing Explorer} ({\rxte}).
Follow-up observations of the ASM and the Proportional Counter Array
(PCA) showed that {\xt} reached its peak intensity of $\sim$2~Crab
on Apr 1.04, well within a day from the onset of the outburst.
On Apr 2.63, \citet{hje98a} found a variable radio source within
the PCA error circle \citep{mar98}.
The position of the radio source coincides
with that of the optical variable star {\ci} \citep{wag98},
which establishes {\ci} being the optical counterpart of {\xt}.

The observations of {\sax} on Apr 3 and 9 and of {\asca} on Apr 3-4
revealed that, unlike the other X-ray transients which harbor
a black hole (BH) or a neutron star (NS) \citep{tan96}, 
the X-ray spectra up to 10~keV have
optically thin thermal nature with a plenty of K-shell emission lines
from highly ionized O, Ne, Si, S, and Fe
\citep{fro98,orr98,ued98}.
The spectra can be represented well
by a two temperature optically thin thermal plasma emission model
with the temperatures of $^{<}_{\sim}$~1~keV and 3--6~keV.
Significant cooling of the plasma and the intensity declination
were found both between the two {\sax}
observations and within the single {\asca} observation.
From a detailed X-ray temporal analysis based on the PCA data 
during the outburst was found no rapid random variability \citep{bel99},
which is remarkably different from the {\ns} and {\bh} transients.
The X-ray to radio light curves \citep{fro98} clearly show that the burst peak
occurs later for longer wavelengths, and the $e$-folding decay time
is shorter for shorter wavelengths.
The Burst and Transient Source Experiment ({\batse}) on board
{\it Compton Gamma Ray Observatory} ({\cgro}) recorded the shortest
timescales among all of the wavebands, 
the intensity arriving at the peak within only $\sim$~0.1~d and 
its $e$-folding decay time being $\sim$~0.56~d \citep{har98}.
The hard X-ray spectrum in the band 20--100~keV obtained with
the {\batse} between Apr. 1.0--2.0 is represented by a power law
with a photon index of $3.9\pm 0.3$.
The flux is compatible with
that of the contemporaneous {\rxte} PCA observations
in the band 2--25~keV \citep{bel99}.
In the optical spectra, a broad blue wing with the velocity of more than
2500~km~s$^{-1}$ was detected from hydrogen Balmar emission lines
only during the outburst \citep{rob02,hyn02}.
By comparing the radio data on Apr 2.63 and 3.83,
\citet{hje98b} found that the radio spectra showed a transition
from optically thick to thin.
The results described described above indicate that the radiation from radio to
$\gamma$-ray during the outburst originates from an expanding gas
ejected by a sort of eruptive event.

Observations of {\xt} in quiescent state, on the other hand, 
were carried out by {\sax} in 1998 September \citep{orl00}, 
1999 September and 2000 February \citep{par00}, and by {\xmm}
in 2001 August \citep{boi02}.
Except for the third {\sax} observation only giving an upper limit
to the absorption-corrected 1--10~keV luminosity
of $< 2.5\times 10^{33}$~erg~s$^{-1}$, 
the other three observations positively detected {\xt} at the luminosities
in the range $(1.4 - 23)\times 10^{33}$~erg~s$^{-1}$.
The X-ray spectrum obtained by {\xmm} is likely to be composed of
two components; one dominates in the band below $\sim$~4~keV with little
absorption, and the other, undergoing heavy photoelectric absorption
with $N_{\rm H} = 5^{+3}_{-2}\times 10^{23}$~cm$^{-2}$,
is conspicuous above $\sim$~5~keV.
It seems that the spectrum of the first {\sax} observation
is dominated by the former component, 
whereas that of the second is so by the latter.
It is worth noting that 
a strong iron emission line centered at $6.43\pm 0.09$~keV is detected
in the {\xmm} observation 
with the equivalent width of $940^{+650}_{-460}$~eV.
An emission line, probably attributable to iron, 
is also detected at $7.0^{+1.6}_{-0.2}$~keV and $7.3\pm 0.2$~keV, 
respectively, from the first and second {\sax} observations
\citep{orl00,par00}.

The optical nature of {\ci} had been a matter of debate until recently.
It was originally designated as MWC~84
in a list of Be stars with infrared excess \citep{all76}
which are later recognized as B[e] stars.
Although a symbiotic characteristic is reported \citep{ber95},
\citet{bel99} argued against this based on their infrared to optical spectrum,
and claimed it to be classified as a B[e] star.
Taking into account the high bolometric luminosity
and composition of the optical emission lines,
\citet{rob02} finally categorized {\ci} into a supergiant B[e] (sgB[e]) star,
which has prodigious mass loss rate $> 10^{-5} M_\odot$~yr$^{-1}$
\citep{dfr98} and high bolometric luminosity 
$10^5 < L_{\rm bol}/L_\odot < 10^6$ \citep{zic98}.

The distance to {\ci} had been estimated to be $1-2$~kpc based on
the luminosity, and the extinction-distance relation
\citep{zor98,bel99,cla00,orl00}.
\citet{rob02}, however, pointed out that the extinction-distance
relation is unreliable in the direction of {\ci}, because the interstellar
matter is patchy.
They proposed a new distance estimate based on 
the velocity of optical emission lines with the aid of the Galactic rotation
model \citep{bur88a}, distribution of matter in the Galactic disk 
\citep{bur88b}, and distances to the known \ion{H}{2} region 
\citep{bli82,cha95}.
\citet{hyn02} also estimated the distance
by making use of a velocity structure of the interstellar Na D absorption line.
The authors of these two papers arrived at almost the same conclusion
that the distance to {\ci} is in the range 5--9~kpc.
We hereafter adopt the distance of 5~kpc as a default according to
recent convention.

One of the most controversial issues as yet unsettled
for {\ci} is the identification of the accreting compact object.
There are, of course, three possibilities: either a {\bh},
a {\ns}, or a white dwarf (WD).
\citet{orl00} pointed out that the outburst behavior of {\ci} is similar to 
nova outburst, and that the compact object might be a {\whd}.
\citet{bel99} argued for a {\ns} since the
overall outburst behavior is similar to that of the 69~ms X-ray pulsar
A0538$-$66 in LMC \citep{cor97}.
\citet{rob02} obtained the X-ray luminosity of
$L(2-25\mbox{ keV}) = 3\times 10^{38}$~erg~s$^{-1}$ based on their revised
distance of 5~kpc, making {\ci} one of the most luminous X-ray transient.
Comparing the outburst luminosity to that during quiescence,
they concluded that the compact object is most likely a {\bh}.

In this paper, we present results from {\asca} observations of {\ci}
in quiescence and those from more detailed analysis of the soft
X-ray component in outburst dominating below $\sim$~1~keV \citep{ued98},
which is probably identical with that detected in
one of the 1998 Apr observations by {\sax} \citep{orr98}.
Overall X-ray behavior revealed by the {\asca} observations 
is consistent with the picture that the compact object in {\ci} is a {\whd},
and the eruptive event that triggers the hard X-ray outburst
can be identified with a nova outburst
(= thermonuclear flash on the surface of the {\whd}).
We note that the errors quoted throughout this paper are those
at the 90~\% confidence level, unless mentioned otherwise.

\section{Observations}

The {\asca} observation of {\ci} in quiescence was performed
during 1999 Feb 19.415-20.445,
which was between the first (1998 Sep 3-4) and second (1999 Sep 23-35) 
{\sax} observations of {\ci} in quiescence \citep{orl00,par00}.
Throughout the observation,
Solid-state Imaging Spectrometer (SIS; \citet{bur94}; \citet{yam99})
was operated in 1-CCD Faint mode, and 
Gas Imaging Spectrometer (GIS; \citet{oha96}; \citet{max96})
was so in PH mode with the default telemetry bit assignment.
We retrieved the data from the Data ARchive and Transmission System (DARTS)
\footnote{http://www.darts.isas.ac.jp/} operated in the Institute of
Space and Astronautical Science (ISAS).

We applied the following selection criteria to both the SIS and GIS data.
We did not use the data while the spacecraft locates within 60~s from
the South Atlantic Anomaly.
We also discarded the data while the pointing direction of the telescope
was within 5~deg from the Earth limb.
In addition to these, we further adopted the following two criteria only
for the SIS.
We discarded the data while the pointing direction of the telescope
was within 20~deg from the edge of the bright Earth limb
illuminated by the Sun.
After these criteria being applied, 
some 37.4~ks and 43.8~ks remain as the good time intervals
for the SIS and the GIS, respectively.
For source photon integration regions, we adopted a circle
with a radius of 2~arcmin and 3~arcmin centered on {\ci} for the SIS
and the GIS, respectively.
For background regions, we used the residual area of the same CCD chip
employed for the {\ci} observation that is out of the source integration
region for the SIS, while, for the GIS, we adopted a circle with a
radius of 6~arcmin whose center locates opposite to the source integration
region with respect to the optical axis of the telescope, in order to
take vignetting of the telescope into account.
After background subtraction, the average count rates throughout the
observation are $0.054 \pm 0.003$ and $0.058 \pm 0.004$~c~s$^{-1}$
per detector for the SIS in the band 0.5--10~keV and
the GIS in the band 0.7--10~keV, respectively.

The {\asca} observation of {\ci} in outburst was carried out
from 1998 Apr 3.31 to 4.14, which was only three days after the onset
of the outburst, and only two days after the X-ray peak \citep{smi98}.
Detailed observation journal was described in \citet{ued98}.
The spectrum was represented by the two temperature optically thin thermal
emission model with temperatures of 1.1~keV and 5.7~keV
with relative abundances of Si, S, and the other metals
being 1.25$\pm$0.04, 1.03$\pm$0.13, and 0.36$\pm$0.02, respectively,
relative to the cosmic composition \citep{angr89}.
A neutral iron emission line at 6.41~keV was found with an equivalent width
of 90~eV.
The entire spectrum was covered by neutral absorber with 
$N_{\rm H} = (4.6\pm 0.3) \times 10^{21}$~cm$^{-2}$.
In addition to this hard X-ray component, they also discovered the
other spectral component
which appears and flickers in the latter half of the observation.
Its spectrum has a sharp cutoff at $\sim 0.8$~keV, and only detected
below 0.9~keV \citep{ued98}.
They could fit the spectrum with a blackbody model
with $kT = 0.12\pm 0.02$~keV with K-edges of hydrogenic and He-like
oxygen at $\sim$0.77~keV and $\sim$0.84~keV.
We are interested in this soft component, which is not analyzed
in full detail by \citet{ued98}.
Accordingly, we concentrate on the SIS in analyzing the outburst data,
because the SIS has higher sensitivity below 1~keV than the GIS.
The same selection criteria as those for the quiescence data are applied,
except that the radius of the source photon integration region
is taken to be 3~arcmin.

\section{Analysis and Results}

\subsection{Light Curve in Quiescence}

We have first made light curves of {\ci} in quiescence, in order to see
if there is any variability during the {\asca} observation.
According to the data selection criteria (\S~2),
we have extracted source and background photons for each
detector (SIS0, SIS1, GIS2, GIS3) separately.
With a bin width of 6400~s, 
we have created light curves of the SIS by adding up the photons
from the two SIS detectors after background subtraction.
The same procedure is applied for the GIS.
The large bin size of 6400~s is adopted
because of statistical limitation.
As shown in the next subsection,
the {\asca} spectrum (Figure~\ref{fig:ave_spec_qui})
and also the {\xmm} spectrum \citep{boi02} both in quiescence
obviously consist of two spectral components separated
at around a few keV.
Accordingly, the light curves are made in the two bands 0.5--2.0~keV (soft) and
2.0--7.0~keV (hard) as well as 0.5--7.0~keV (total) for the SIS,
and 0.7--2.0~keV (soft),
2.0--7.0~keV (hard) and 0.7--7.0~keV (total) for the GIS.
In order to improve statistics, we have further summed up 
the SIS and GIS light curves in the corresponding bands,
and then have fitted them with a constant model.
The resultant $\chi^2_\nu $~(d.o.f.) is 
0.65~(13), 1.10~(13), and 1.46~(13) for the soft, hard, and total light curves,
respectively.
Although one can find that both the soft and hard components are variable 
in a longer timescale 
by comparing the three {\sax} observations \citep{orl00, par00} and
the observation of {\xmm} \citep{boi02}, 
their fluxes are consistent with being constant within the
single {\asca} observation on a timescale between $\sim$2~h and 1~d.

\subsection{Spectrum in Quiescence}

\begin{figure}
\rotate
\epsscale{0.8}
\plotone{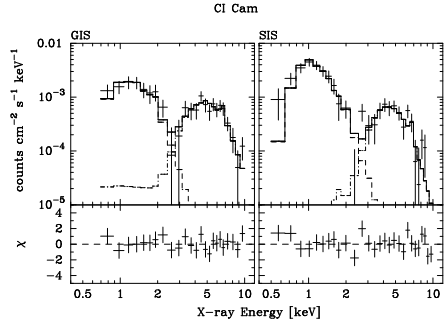}
%\plotone{fig1.eps}
\caption{
The averaged GIS and SIS spectra of {\ci} during quiescence, together with
the best-fit 2 {\mekal} model with the common abundance of 0.36
times the cosmic \citep{angr89}.
For the other parameter values, see the ``2 {\mekal} III'' column in Table~2.
The spectra of the GIS and the SIS are fitted simultaneously.
\label{fig:ave_spec_qui}
}
\end{figure}

Since there is no evidence for variability,
we evaluate the spectra averaged throughout the observation,
which are shown in Figure~\ref{fig:ave_spec_qui}
for the SIS and the GIS separately.
\placefigure{fig:spec_sis}
As obviously seen,
they are composed of the two continuum components separated at 2--3~keV.
Accordingly, we have attempted to fit them
with soft and hard components undergoing independent
photoelectric absorptions.
For spectral fitting, we use XSPEC v11.2 \citep{arn96}.
Throughout this subsection, the GIS and SIS spectra are always fitted
contemporaneously.

First of all, we have attempted to fix spectral models for
both emission components from a purely statistical point of view.
We have tried optically thin thermal plasma emission ({\mekal}),
power law, and blackbody models for the soft component, and
{\mekal}, power law, and thermal bremsstrahlung models for the hard component.
As noticed from Figure~\ref{fig:ave_spec_qui}, the {\asca} spectra suggest
presence of an iron {\ka} emission line feature around 6-7 keV.
The iron {\ka} emission lines have also been detected
from the {\sax} and {\xmm} observations \citep{par00,boi02}.
We thus have added a narrow Gaussian, with its central energy free to vary,
in the case that either the power law or the thermal bremsstrahlung is
adopted for the hard component.
In order to check the validity of adding the Gaussian line,
we have tried a model composed of a soft {\mekal} and a hard power law,
and by including a Gaussian, we have found that
the resultant $\chi^2$ value is improved from 38.23 (44 d.o.f.) to
31.78 (42 d.o.f.), implying $\Delta\chi^2 = 6.46$ for 2 degrees of freedom.
According to F-test, the inclusion of the iron line into the {\asca} spectra
is justified at the $98.0$~\% confidence level.

The best-fit $\chi^2_\nu$ matrix for the fits
with the trial models is summarized in Table~\ref{tab:softhard}.
\placetable{tab:softhard}
There is no reason to prefer any particular model statistically
for both soft and hard components.
We thus need some physical consideration in choosing a particular pair of
the models.
It has been suggested that the soft component probably originates from
the sgB[e] star \citep{rob02}.
The temperature of $\sim 0.22$~keV \citep{orl00} and the 0.5-2.0~keV
luminosity of $2\times 10^{34}$~erg~s$^{-1}$ is consistent with
the X-ray emission from sgB[e] star \citep{rob02}.
The power-law fit to the soft band spectrum of the current {\asca} data
results in the photon index of $5.3^{+1.8}_{-1.1}$.
Such a steep spectrum probably corresponds to a high energy end of
a thermal spectrum.
Accordingly, we hereafter adopt the {\mekal} model for the soft component.

For the hard component, on the other hand, since the {\mekal} model and
the thermal bremsstrahlung plus Gaussian model are essentially the same,
we compare the power-law model (plus Gaussian) with the {\mekal} model.
The results of the fits in the band 0.5-10~keV, including the soft {\sc mekal}
below $\sim 2$~keV, are summarized in Table~\ref{tab:fit_qui}.
\placetable{tab:fit_qui}
For the {\mekal} components,
we have first tried to set their abundances free to vary independently.
(``2 {\mekal} I'' model in Table~\ref{tab:fit_qui}), and have found that
they become consistent within the errors.
We thus have constrained the two abundances to be the same
(``2 {\mekal} II'' model).
The fit with this model results in the abundance to be $0.26^{+0.54}_{-0.26}$
times the cosmic value \citep{angr89}.
The error is, however, large due to statistical limitation.
We thus have fixed them at the value 0.36 times the cosmic
obtained from the outburst observation by \citet{ued98}
(``2 {\mekal} III'' model).
As can be seen from Table~\ref{tab:fit_qui}, all of the three {\sc mekal} fits 
are acceptable at the 90~\% confidence level, 
and the fit with ``2 {\sc mekal} I'' model is formally the best.
As described above, however, the soft component can be interpreted as the 
coronal emission from the sgB[e] star, and 
since the hard component emission is probably a result of mass accretion
from the sgB[e] star, there is no reason that
the abundance of the hard {\sc mekal} component is different from
that of the soft.
Accordingly, we believe the last ``2 {\mekal} III'' model is physically
most reasonable.
Based on this model, the best-fit temperatures of the soft and hard components
are obtained to be $kT_1 = 0.45^{+0.17}_{-0.14}$~keV
and $kT_2 = 5.5^{+7.7}_{-2.9}$~keV, respectively.
Bolometric luminosities of the hard {\mekal} component is calculated from
the emissivity of thermal bremsstrahlung \citep{ryb79}
at the observed temperature and the emission measure.
Those of the soft {\mekal}, on the other hand, is calculated 
from the emissivity of the optically thin thermal plasma \citep{gae83}.
Around the plasma temperature of $0.4-0.5$~keV observed, the emissivity of
the plasma is dominated by the iron L line emissions.
Accordingly, we have reflected the observed iron abundance
into the luminosity calculation.
As a result, the bolometric luminosities of the soft and hard components
based on ``2 {\mekal} III'' model are obtained to be
$6.0 \times 10^{33}$~erg~s$^{-1}$ and $9.4 \times 10^{33}$~erg~s$^{-1}$,
respectively.

The power-law model, on the other hand, can fit the spectra equally well
as the {\sc mekal} models.
We thus next consider the non-thermal power law as the continuum emission model
of {\ci} in quiescence.
In this case, the observed iron emission line should be of fluorescence origin.
We assume that the line originates from matter
surrounding the hard component emitter uniformly with the $4\pi$ solid angle.
The thickness of the matter is then equal to
the line of sight photoelectric absorption
$N_{\rm H2}=1.9^{+0.4}_{-0.2}\times 10^{23}$~cm$^{-2}$
(Table~\ref{tab:fit_qui}).
From this hydrogen column density, the equivalent width of the iron {\ka}
emission line is expected to be $190^{+40}_{-20}$~eV under the condition
that the intrinsic photon index of a power-law emission is 1.1
and that the fluorescing matter has the cosmic composition \citep{ino85}.
In the case of {\ci}, there are two factors to reduce the expected equivalent
width; one is the iron abundance,
which is 0.36$\pm$0.02 times the cosmic \citep{ued98},
that reduces the equivalent width proportionally;
the other is the larger photon index of 2.28 (Table~\ref{tab:fit_qui}),
which also reduces the equivalent width according to the following
scaling relation;
\[
EW(\gamma)\; \propto \; \frac{1}{f(E_{\rm K\alpha}; \gamma )}
\int^{\infty}_{E_{\rm edge}}f(E; \gamma)\; \sigma(E; E_{\rm edge})\; dE
\]
\citep{ezu99}, where $f(E_{\rm K\alpha}; \gamma )$ represents
the intrinsic X-ray photon flux density of the power law
with the photon index $\gamma$ at the energy $E_{\rm K\alpha}$,
and $E_{\rm K\alpha}$ and $E_{\rm edge}$
are the energies of the {\ka} line and the K-edge of the neutral iron,
which are 6.40~keV and 7.11~keV, respectively.
$\sigma$ is the cross section of the iron K-edge photoelectric absorption.
In the case of $\gamma = 2.28$, we have obtained the $EW$-scaling factor
from the $\gamma = 1.1$ case to be 0.64,
assuming $\sigma(E; E_{\rm edge}) \propto E^{-3}$ for $E > E_{\rm edge}$.
As a result, the expected equivalent width from {\ci}
becomes only $\sim$40~eV with the 90\%upper limit of $80$~eV
within the errors of the related parameters.
This cannot be compatible with the observed equivalent width
of $624^{+418}_{-495}$~eV (Table~\ref{tab:fit_qui}).
It seems that the observed iron {\ka} emission line is difficult to
be explained solely by the fluorescence,
and that the underlying continuum is likely to be 
the optically thin thermal plasma emission, supplying additional
iron {\ka} line emission.
In Figure~\ref{fig:ave_spec_qui},
we show the best-fit ``2 {\mekal} III'' model by histograms, 
which we put forward as the best representation of the {\ci} spectra.
We note, however, that this conclusion is derived under the hypothesis
of the fluorescing matter surrounding the central object uniformly.
This may, however, not be true, because there is a dense circumstellar disk
around the sgB[e] star out of the line of sight.
Also the absorber is assumed to be neutral, but this may not be true, either.
Both these effects provide with a larger equivalent width
at a given line of sight column density estimated with neutral matter
cross section.
Hence, we cannot rule out the power-law model
for the hard spectral component of {\ci} in quiescence.

Note also that,
although we have shown that the thermal iron {\ka} line is preferred,
partial contribution of the fluorescent component is not ruled out.
In the {\xmm} observation \citep{boi02}, 
the iron line is detected at the energy $6.43\pm 0.09$~keV
with the equivalent width of $940^{+650}_{-460}$~eV,
which is greater than that obtained with {\asca}.
Since the line of sight hydrogen column density during the {\xmm} observation
is also greater than during the {\asca} observation,
the greater equivalent width and the lower line central energy
are probably brought about by the increment
of the contribution of the fluorescent component.
The upper limit of the line width 0.28~keV obtained by {\xmm}
probably allows contribution from the thermal plasma component
to the observed iron line.
Hence, the iron {\ka} emission line from {\ci} is probably
a mixture of the thermal plasma and fluorescent components.
For the {\asca} spectra in Figure~\ref{fig:ave_spec_qui},
it is possible to insert an additional Gaussian 
representing the fluorescent iron line into the model,
which we have, however, refrained from because of statistical limitation.

\subsection{Reanalysis of the Soft Component in the Outburst Spectrum}

\subsubsection{Light curve}

\begin{figure}
\epsscale{0.8}
\plotone{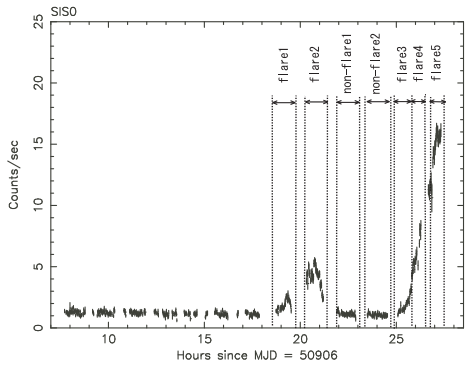}
%\plotone{outburst_lcurve.eps}
\caption{
The SIS0 light curve of {\ci} in the band 0.5--1.0~keV during the outburst.
\label{fig:lc_outb}
}
\end{figure}

As described in \S~1, the X-ray emission from {\ci} during the outburst
is dominated by that from the optically thin thermal plasma with a temperature
of several keV \citep{ued98}, 
which is probably escaping from the binary system.
In addition to this, a separate soft X-ray component
that is detected only below 1~keV
starts to appear in the middle of the {\asca} observation.
Although it once fades away, it brightens again nearly the end of the
observation.
We show the SIS0 light curve during the outburst in Figure~\ref{fig:lc_outb}.
\placefigure{fig:lc_outb}
As noted by \citet{ued98}, the satellite telemetry is saturated
after around Apr. 4 2:40 due to the increasing intensity of the soft component,
and the detection efficiency of the SIS at the image core is significantly
reduced as a result of event pile-up.
We have corrected this effect by utilizing photons well outside the overflowed
core and a saturated region in drawing the light curve.

\begin{figure}
\epsscale{0.8}
\plotone{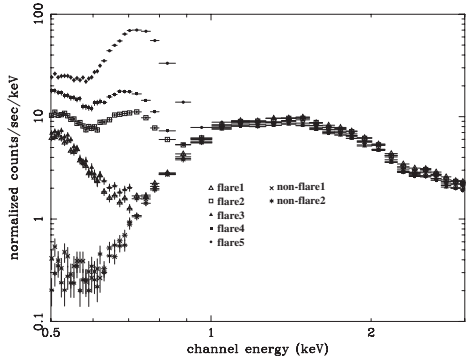}
%\plotone{phsspec2.eps}
\caption{
The SIS0+SIS1 spectra of {\ci} from the segments ``flare 1--5'' and
``non-flare 1 \& 2'' in Figure~\ref{fig:lc_outb}.
\label{fig:phsspec}
}
\end{figure}

In order to examine the nature of the soft component, we have divided 
the full dataset into several pieces 
designated as ``flare 1~\&~2'', ``non-flare 1 \&2'' and ``flare 3--5''
in Figure~\ref{fig:lc_outb}, as delineated by the dashed vertical lines,
and have drawn the average SIS spectra during those intervals separately,
which are shown in Figure~\ref{fig:phsspec}.
\placefigure{fig:phsspec}
We have corrected the event pile-up effect for the spectrum 
during the ``flare 5'' interval.
As obviously seen from the spectral variation, what changes during 
the soft flare is not the line of sight absorption 
but the emission component itself;
the spectrum extends to a higher energy band as the intensity increases.
Owing to a very slow decay of the hard component, the flux levels 
of all of the data segments are similar above $\sim$1~keV.
Accordingly, we regard the average of ``non-flare 1 \& 2'' spectra as the
background and simply subtracted them from the ``flare'' spectra.

\subsubsection{Evaluation of the Spectra by a Power Law}

We have first tried to fit all five flare spectra
after the background subtraction (Figure~\ref{fig:soft_bestfit})
by a power law undergoing the photoelectric absorption.
We have used the spectral energy bins in the band 0.5--0.7~keV 
for ``flare 1 \& 3'', and 0.5--0.9~keV for the others.
The fits are, however, generally poor.
\citet{ued98} pointed out that the high energy cutoff
starting at $\sim$0.8~keV seen in the ``flare 5'' spectrum
is as steep as what can only be realized by a sharp atomic edge.
According to them, we have overlaid the power law with two edges,
and have tried to fit the spectra again.
In doing this, we have fixed the hydrogen column density at
$4.6\times 10^{21}{\rm cm}^{-2}$,
which is obtained by the fit to the hard X-ray component during outburst
out of the soft flare \citep{ued98}.
The results are summarized in Table~\ref{tab:fit_pl_edges}.
\placetable{tab:fit_pl_edges}
The fits for all five spectra are markedly improved and now
acceptable at the 90~\% confidence level.
Although the power-law model is acceptable as an underlying continuum,
however, it is difficult to understand the large variation of the photon index 
ranging 0.5--5.3 over the five flare phases.
No power-law X-ray source has been known so far 
whose photon index varies in such a wide range on the timescale of a day.
The apparent large photon index variation thus probably represents
a temperature variation of the continuum.
We therefore examine a blackbody model in the following subsection.

\subsubsection{Evaluation of the Spectra by a Blackbody}

We have attempted to fit all of the 5 flare spectra
with the blackbody model in the same energy bands as
in the previous subsection.
As we already know, some absorption edges are necessary to fit the spectra.
We thus have introduced the absorption edges one by one
until we obtain acceptable $\chi^2$ values.
The results are summarized in Table~\ref{tab:fit_NHfree}.
\placetable{tab:fit_NHfree}
As suggested by Figure~\ref{fig:phsspec}, the blackbody temperature
increases with the intensity whereas no $N_{\rm H}$ decrease
is found with increasing intensity.
Two edges are necessary to fit
the ``flare 2, 4 \& 5'' spectra, one edge for the ``flare 1'' spectrum,
and no edge for the ``flare 3 '' spectrum.
The fit to the ``flare 3'' spectrum is, however, further improved significantly
if we overlay the blackbody with an edge at 0.67~keV, as listed in
the last column of Table~\ref{tab:fit_NHfree}.
Note that the obtained edge energies in the ``flare 1 \& 3'' spectra 
(0.62~keV and 0.67~keV, respectively) are
somewhat apart from those of \ion{O}{7} and \ion{O}{8} K-edges.
Obviously 
we need a more systematic and sophisticated treatment for the oxygen edges,
which is essential to determine the intrinsic
spectral shape and the luminosity of the soft component.

\begin{figure}
\rotate
\epsscale{0.8}
\plotone{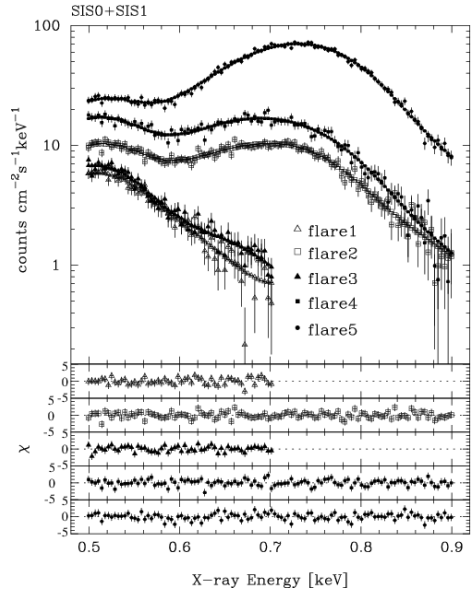}
%\plotone{fit3.eps}
%\plotone{flare_spec_v.eps}
\caption{
The best-fit SIS0+SIS1 spectra of {\ci} from the segments ``flare 1--5''
fitted by the blackbody with the five oxygen edges.
The top panel shows the data and the models for the five ``flare'' spectra.
The other five panels below show 
the fit residuals for the five spectra separately.
For the best-fit parameters, see Table~\ref{tab:fit_NHfix}.
\label{fig:soft_bestfit}
}
\end{figure}

In the second column of Table~\ref{tab:fit_NHfix}, we have tabulated
the energies of ionized oxygen edges above 0.60~keV \citep{ver95}.
It is likely that the edges found in the ``flare 1 \& 3'' spectra are
associated with \ion{O}{4} or \ion{O}{5}.
We thus have introduced a model composed of a blackbody with five
edges corresponding to \ion{O}{4} through \ion{O}{8}.
If we allow all of their energies to vary independently, however,
they never converge because of the limited energy resolution and the narrow
energy bands available.
We have therefore linked the energies 
$E_{\rm edge\,2}$ through $E_{\rm edge\,5}$ to $E_{\rm edge\,1}$
so that their ratios to $E_{\rm edge\,1}$ are
consistent with the theoretical values
listed in the second column of Table~\ref{tab:fit_NHfix}.
The results of the best-fit to each ``flare'' spectrum are shown in
Figure~\ref{fig:soft_bestfit}, whose parameters are
listed from the third column in Table~\ref{tab:fit_NHfix},
where $\tau_1$ through $\tau_5$ are the optical depths of the edges
at their threshold energies.
\placefigure{fig:soft_bestfit}
\placetable{tab:fit_NHfix}
We have multiplied the model by the photoelectric absorption
of the cosmic composition
with the fixed hydrogen column density of $4.6\times 10^{21}{\rm cm}^{-2}$,
which is obtained from the same outburst observation while the soft component
is absent \citep{ued98}.
In addition to this, 
we have also overlaid an additional photoelectric absorption with a free
$N_{\rm H}$, in order to see 
if there is excess intrinsic absorption for the soft component.
Metal abundances of the intrinsic absorber are fixed at the values
obtained from the hard emission component during outburst \citep{ued98},
in which Si, S, and the others are 
1.25, 1.03, and 0.36 relative to the cosmic composition.
Although the statistic of each spectrum is rather limited and the depths
of the edges are not always well constrained,
it is likely from Table~\ref{tab:fit_NHfix}
that oxygen ionization proceeds to a higher level as the flux
increases; the best-fit values of $\tau_4$ and $\tau_5$ are equal to zero
for the low intensity level spectra ``flare 1 \& 3'', nearly all of the
edges are significant for the middle flux spectrum ``flare 2'', 
and only the He-like and hydrogenic edges are significant for the high
flux spectra ``flare 4 \& 5''.
Note that $E_{\rm edge\,5}$ becomes out of the
upper boundary of the fit energy range (0.5--0.9~keV) for some spectra.
This is, however, still significant to cut off the high energy part of
the spectra because of a finite energy resolution of the SIS.
As a result, the blackbody temperature of the soft component
varies in the range $0.07-0.12$~keV.
Although bolometric luminosities are not always well constrained,
they include the Eddington luminosity of a $1M_\odot$ object
for the assumed distance of 5~kpc.

\section{Discussion}

\subsection{Accreting Object Suggested from Data in Quiescence}

We have argued in \S~3.2 that the hard component of the {\ci} spectra
in quiescence is likely to be the optically thin thermal plasma emission
with the temperature of $kT_2 = 5.5^{+7.7}_{-2.9}$~keV,
based on the discussion of the iron line equivalent width.
This type of spectrum reminds us of hard X-ray emission
from accreting {\whd}s (cataclysmic variables).
\citet{muk93} have summarized the {\exo} observations of dwarf novae (DNe)
and have found that the spectra can be represented by a thermal bremsstrahlung
with the temperature of several keV in general,
plus an iron emission line at 6.7~keV.
Their luminosities are, however,
in the range $7\times 10^{30} - 3\times 10^{32}$~erg~s$^{-1}$,
which is smaller than that of {\ci} in quiescence
by more than an order of magnitude.
Note, however, that the mass-donating companion stars in these DNe are
a low-mass main sequence star.
In such a system, mass accretion rate is lower in general than in the system
which includes an evolved companion.
In comparing to {\ci},
we thus have to refer to an accreting {\whd} binary including an evolved
mass-donating star.
One of the best such systems is the symbiotic star {\ch}.
\citet{ezu98} analyzed the {\asca} data of {\ch}
which is the binary composed of a {\whd} 
and a Roche-lobe-filling M6.5 giant star.
We would like to stress, in particular, remarkable similarity
of the {\ch} spectrum (see Figure~2 of \citet{ezu98}) to that of {\ci};
it obviously consists of two components separated at 2~keV;
the harder component is represented by the optically thin thermal spectrum
with the temperature and the metal abundance
of $7.3\pm 0.5$~keV and $0.43\pm 0.05$ cosmic,
undergoing heavy photoelectric absorption with $N_{\rm H}$
as large as $10^{23}$~cm$^{-2}$, which is probably the result of huge amount
of matter supplied by the evolved star,
thereby surrounding the accreting object;
the bolometric luminosity of the hard component is
$1\times 10^{33}$~erg~s$^{-1}$ \citep{ezu98}.
Because of all these similarities to the accreting white dwarf binaries,
a {\whd} is the most likely candidate as the accreting object in {\ci}.

The hard X-ray nature of the {\bh} transients, on the other hand,
has been revealed by recent {\chan} observations.
Their X-ray spectra are represented
by a power law with the photon index of 1.7--2.3 \citep{kon02},
which covers that of {\ci} (= 2.28; Table~\ref{tab:fit_qui}).
Although the X-ray luminosities are found to be mostly 
in the range $10^{30}-10^{32}$~erg~s$^{-1}$ \citep{gar01},
which are significantly lower than that of {\ci}
($\simeq 10^{34}$~erg~s$^{-1}$; Table~\ref{tab:fit_qui}),
the longest orbital period system {\vcyg} (= {\gin}) is found to
have a luminosity as high as $5\times 10^{33}$~erg~s$^{-1}$
\citep{gar01,asa98b}.
The power law with the photon index of $\sim$2 can thus be realized 
at the luminosity level up to $\sim 10^{34}$~erg~s$^{-1}$
in the {\bh} transients that include an evolved mass donor.
Since the power-law model can fit the hard component of {\ci}
in quiescence equally well as the {\sc mekal} model (Table~\ref{tab:fit_qui}),
as described in \S~3.2, we cannot rule out the possibility of a {\bh}
as a candidate of the accreting object of {\ci} from the analysis
of the {\asca} data in quiescence.

Due to the same reason, we cannot exclude a {\ns} possibility, either.
The X-ray spectra of the {\ns} transients in quiescence are represented
well by a blackbody model with a temperature of $0.1-0.3$~keV \citep{asa98b},
or a {\ns} atmosphere model \citep{rut02}, occasionally accompanied
by the power-law that has nearly the same photon index as that found in
the {\bh} transients.
It is thus possible that the hard component of {\ci} in quiescence
is the power-law emission often detected from the {\ns} transients.
\citet{nar96} claim that the power-law component is emission
from the advection-dominated accretion disk (ADAF) formed around the
central object.
If so, the blackbody component from the {\ns} surface in {\ci}
should be obscured by the heavy photoelectric absorption 
($N_{\rm H} \simeq 1.9\times 10^{23}$~cm$^{-2}$; Table~\ref{tab:fit_qui})
and invisible.
The soft component should thus be attributed to the coronal emission of 
the sgB[e] star, as described in \S~3.2.

In summary, because of the similarities of the {\ci} spectra in quiescence
to other mass-accreting {\whd}s, 
we argue that the accreting object of {\ci} is most likely a {\whd}.
This argument owes mostly to the identification of the hard X-ray component
to the optically thin thermal plasma emission, which is, however, not
completely established due to statistical limitation.
The spectral slope of the hard component is similar to
that of the {\bh} and {\ns} transients in quiescence
if we adopt the power law in evaluating the spectra.
Hence, we cannot rule out possibilities of a {\ns} and a {\bh} as 
the candidates of the accreting object in {\ci}.

\subsection{Accreting Object Suggested from Outburst Data}

The updated distance of 5~kpc to {\ci} results in
the hard X-ray luminosity at the burst peak of 
$L(2-25\mbox{ keV}) = 3\times 10^{38}$~erg~s$^{-1}$,
which exceeds the Eddington limit of a $1M_\odot$ star.
On the other hand, the peak luminosities so far obtained from 
other {\ns} and {\bh} transients are, in general,
in the range $10^{37} - 10^{38}{\rm erg\;s}^{-1}$ and
$10^{38} - 10^{39}{\rm erg\;s}^{-1}$
for the {\ns} and {\bh} transients, respectively \citep{tan96,che97}.
The burst peak luminosity of {\ci} thus matches that of the {\bh} transients.
These have been the main reasons to put forth the {\bh} identification of {\ci}
(see \citet{rob02}, for example).

As reported by \citet{cor97}, however, 
the luminosity of the transient X-ray pulsar A0538$-$66 in LMC
reaches as high as $\sim 10^{39}$~erg~s$^{-1}$ at the burst peak, 
which violates both the Eddington limit of the {\ns} and the empirical
source classification according to the burst peak luminosity.
There are some pieces of evidence
that the plasma emitting the hard X-rays during the {\ci} outburst
is not the one accreting steadily onto
the central object but escapes from the binary (see \S~1).
One thus has to be prudent enough to
apply the Eddington limit luminosity to the outburst of {\ci}.
We believe we cannot exclude the {\whd} from the candidates
of the central object of {\ci}
simply because the hard X-ray luminosity at the burst peak
exceeds the Eddington luminosity of the typical $1M_\odot$ object.

Although a nova has been regarded as a soft X-ray emitter,
some novae have been found to show hard X-ray emission
in an early phase of their outbursts.
The earliest detection of the hard X-ray emission from novae
is obtained from the {\ro} PSPC observation of V838~Her
five days past the optical peak \citep{llo92}.
{\asca} detected the hard X-ray emission from the fast nova V382~Vel
twenty days after the optical peak.
The {\asca} spectrum of V382~Vel can be well fitted 
by the thermal bremsstrahlung model
with the temperature of $\sim 10$~keV \citep{muk01}.
The optically thin nature and a significant X-ray flux over 10~keV
are similar to the {\asca} spectrum of {\ci} obtained 3 days after
the onset of the outburst \citep{ued98}.
We remark that no soft X-ray transient that harbors
a {\ns} or a {\bh} shows the optically thin thermal nature 
in the X-ray spectrum during its outburst.

In addition to this,
remarkable similarity of the soft component to that of the super-soft source 
(SSS) further strengthens the {\whd} identification of {\ci}.
The spectrum of one of the SSS's CAL~87 obtained by {\asca} 
is well represented by the blackbody model with $kT_{\rm bb} = 75$~eV
multiplied by the K-edges of \ion{O}{7} and \ion{O}{8}, 
and the bolometric luminosity varies in the range of $4\times 10^{37}$ 
to $1.2 \times 10^{38}$~erg~s$^{-1}$ \citep{asa98a,ebi01},
which are both common to those of the {\ci} soft component.
It has been pointed out that the nova can show properties of the SSS,
after the photosphere shrinks back onto the {\whd} surface in a declining
phase of the outburst.
\citet{sta00} discovered a super-soft continuum
underneath a line-rich spectrum of the nova V1494~Aql 
by the {\chan} grating observations.
The super-soft continuum is also found from V382~Vel by {\sax} after 6 months
\citep{ori02},
and V1974~Cyg (= Nova Cyg 1992) by the {\ro} PSPC 
after 8 months \citep{kra96} from their outbursts.
The soft component of {\ci} detected by {\asca} can thus
be ascribed most naturally to the super-soft X-ray emission
from the {\whd}.
It is possible to interpret its on-and-off behavior
manifested in Figure~\ref{fig:lc_outb} by repeated bounces of
the {\whd} photosphere, which is subject to the radiation pressure
in high luminosity environment close to the Eddington limit.

\subsection{Distance Estimate Based on Nova Hypothesis}

In the preceding two subsections, we have shown some pieces of evidence 
indicating the central object of {\ci} being the {\whd},
based on the quiescence and outburst observations,
and have proposed a nova outburst picture for the {\ci} outburst.
It is well known that the nova outburst is a good distance indicator,
since there is a tight correlation between the absolute
peak visual magnitude ($M_V^{\rm max}$) and the speed class \citep{war95},
the latter of which is characterized by the elapsed time
for the nova to decline by 2 and 3 magnitudes from the peak,
which are conventionally denoted as $t_2$ and $t_3$, respectively.
As described in \S~1, the distance to {\ci} is estimated to be $5-9$~kpc.
Although the characteristics of the nova light curve could be significantly
modified with the circumstellar environment of {\ci},
it seems still worthwhile to estimate the distance to {\ci}
under the assumption of the nova outburst.

\begin{figure}
\rotate
\epsscale{0.8}
\plotone{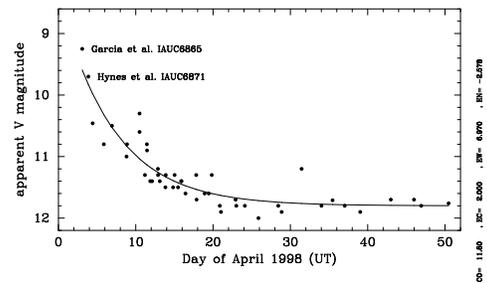}
%\plotone{vsnet.eps}
\caption{
The optical V-band light curve of {\ci} during 1998 outburst.
The first two datapoints are taken from IAUCs.
The others are from VSNET database 
(http://vsnet.kusastro.kyoto-u.ac.jp/vsnet/LCs/index/index.html).
The solid line is the best-fit ``constant + exponential'' model
with the peak of the outburst set at April 2.0 (UT).
For parameter values, see Table~\ref{tab:vsnet}.
\label{fig:vsnet}
}
\end{figure}

The relation between $M_V^{\rm max}$ and $t_2$ is calibrated by \citet{coh88}
as
\begin{equation}
\label{eq:cohen}
M_V^{\rm max} \;=\; 2.41\,(\pm 0.23)\,\log_{10}t_2 \,-\,10.70\,(\pm 0.30),
\end{equation}
where the unit of $t_2$ is a day.
In order to estimate $t_2$, we have retrieved the optical light curve
between Apr.~3 and May.~20 from VSNET database
\footnote{http://vsnet.kusastro.kyoto-u.ac.jp/vsnet/LCs/index/index.html},
which is shown in Figure~\ref{fig:vsnet}, together with some
data points obtained in the earliest phase \citep{gar98,hyn98}.
\placefigure{fig:vsnet}
We have attempted to fit the following model to this light curve;
\begin{equation}
\label{eq:vsnet}
m_V\,(t) \;=\; m_V^{\rm const} - m_V^{\rm amp}\,\exp\,\left( 
	- \frac{t - t_{\rm max}}{\tau_{\rm fold}} \right),
\end{equation}
where $t_{\rm max}$ is the time of the peak magnitude,
$m_V^{\rm const}$ is the magnitude of the sgB[e] star,
and $m_V^{\rm amp}$ is the decrement of the magnitude at the burst peak.
Unfortunately, the earliest optical observation was made on
April 3.08-3.17 \citep{gar98}, at which the optical flux
has probably already started to decline \citep{hyn98}.
The multi waveband light curves \citep{fro98}, on the other hand, indicate
that the burst peak occurs later for the longer wavelength.
It is thus reasonable to assume that the optical peak occurs no earlier
than that of the X-ray, which was April 1.04 \citep{smi98}.
We thus have tried to fit the data with eq.~(\ref{eq:vsnet})
by fixing $t_{\rm max}$ either at April 1.0, 2.0 and 3.0.
The results are summarized in Table~\ref{tab:vsnet}.
\placetable{tab:vsnet}
In obtaining $t_2$, we need to care about contribution of the optical flux
from the sgB[e] star which is completely negligible for ordinary novae.
We thus once convert eq.~(2) to a flux, subtract the flux of the sgB[e] 
star ($\propto 10^{-0.4\,m_V^{\rm const}}$), and then calculate $t_2$
as the time at which the subtracted flux becomes $10^{-0.8}$ of the peak.
Resultant $t_2$ is in the range 5.8--7.2~d depending upon the choice
of $t_{\rm max}$.
Accordingly, {\ci} is classified into Very fast nova ($t_2 < $ 10~d).
The apparent $V$-magnitudes at the peak ($= m_V^{\rm const} - m_V^{\rm amp}$)
are subject to the correction of extinction $A_V = 2.0-4.4$ \citep{hyn02},
resulting in the distance modulus with $M_V^{\rm max}$ obtained
through eq.~(\ref{eq:cohen}) using $t_2$.
It is obvious that most of the error in the distance originates from $A_V$.
We thus neglected all of the other errors such as those of the parameters
in eq.~(\ref{eq:vsnet}) and of the data points in the light curve.
As a result, the distance to {\ci} is estimated to be in the range
5-17~kpc, irrespective of the choice of $t_{\rm max}$ within Apr.~1.0-3.0.
This result is consistent with the estimation 5-9~kpc \citep{rob02,hyn02},
which provides with another support to the picture
that the central object of {\ci} is the {\whd}
and the outburst is ascribed to a nova outburst.

\subsection{A Few Remarks on the White Dwarf Interpretation}

We have so far argued for the picture that the central object of {\ci}
is the {\whd} and its outburst is consistently interpreted as
the nova outburst.
A nova, however, brightens by 7--16 magnitudes in optical
in general by the outburst \citep{war95}.
Compared to this, the optical brightening of {\ci} ($\Delta m_{V} \simeq 2-3$)
is remarkably small.
This apparent inconsistency is due to a high persistent optical flux
from the sgB[e] companion star.
According to \citet{hyn02},
its optical luminosity is as great as $>10^{5.4}L_\odot$,
which implies that {\ci} system is more luminous than the standard nova 
by more than 13 magnitudes in quiescence.
The apparent small brightening of {\ci} is thus caused by 
this high ``background'' level.

Finally, we confess our discomfort to propose
a binary composed of a {\whd} and an OB star for {\ci},
because the existence of such a binary
has never been confirmed observationally,
although it can be formed through mass exchange between
the components, or tidal capture of a {\whd} by an OB star.
We, however, point out that $\gamma$-Cas has been a candidate of such a binary
\citep{kub98,owe99}.
Based on the {\ro} all-sky survey,
\citet{mot97} have found some OB stars show excess X-ray emission above
the level expected from the empirical ratio $L_X/L_{\rm opt}$.
They suggest that some of them may harbor a {\whd}.

\section{Conclusion}

We have presented the results from the {\asca} observations of
{\ci} both in quiescence and in outburst.
The quiescent spectrum of {\ci} consists of the two spectral components
separated at $\sim$2--3~keV.
Among them, the hard component, which undergoes photoelectric
absorption with $N_{\rm H} = 1.9\times 10^{23}{\rm cm}^{-2}$,
is likely to be the optically thin thermal plasma emission,
based on a quantitative discussion
of the iron line equivalent width $624^{+418}_{-495}$~eV.
If so, the central accreting object of {\ci} is probably a {\whd},
because the optically thin thermal nature of the hard X-ray emission
is common among the accreting white dwarf binary, such as cataclysmic variables,
whereas no {\ns} and {\bh} transient in quiescence
manifests such characteristic in its X-ray spectrum.
The spectrum of the hard component can, however, also be fitted
with a power law equally well, which is a common characteristic among
soft X-ray transients in quiescence.
Consequently, possibilities of a {\ns} and a {\bh} cannot be ruled out
solely from the {\asca} data in quiescence.

The outburst data, on the other hand,
obviously favor the picture that {\ci} hosting a {\whd}
due to the following reasons;
\begin{description}
\item[(1)] The spectrum of the soft component intermittently visible 
in the {\asca} observing window during the outburst can be represented well by
the blackbody multiplied by the highly ionized oxygen edges.
This, together with the luminosity as high as
$1\times 10^{38}{\rm erg\;s}^{-1}$, reminds us of the super-soft source CAL~87,
suggesting strongly that {\ci} harbors a {\whd}.
Since some novae are reported to mimic the SSS
during its declining phase, it is possible that the outburst of {\ci} is
a nova outburst.
\item[(2)] Recent observations revealed that
the nova can emit hard X-ray emission
over the energy 10~keV in the form of the optically thin thermal emission
in an early phase of its outburst.
This optically thin thermal plasma emission is also detected from {\ci}
during its outburst,
whereas the optically thin thermal nature has not been found from
any {\ns} or {\bh} transient during the outburst so far.
\item[(3)] By assuming the nova outburst, we can estimate the distance to
{\ci} by means of the decay time constant of the optical light curve.
Within the uncertainty of the burst onset date,
the resulting distance becomes in the range 5--17~kpc,
which strengthens the identification of {\ci} to the nova because
this range matches the updated distance 5--9~kpc to {\ci}.
\end{description}
Synthesizing all results
obtained from the quiescence and outburst observations by {\asca},
we are led to conclude that the central accreting object of {\ci} is a {\whd},
and its outburst can be regarded as a nova outburst.

\acknowledgments
We are grateful to VSNET administrators and participants 
for their effort to accumulate data of {\ci} and make them
open for public use.
We would like to express our special thanks to Dr. T. Dotani (ISAS) for his
useful discussions,
and the anonymous referee for his/her constructive comments.

%%%%%%%%%%%%%%%%%%%%%%%%
\begin{deluxetable}{rccc}
\tablecolumns{4}
\tablewidth{0pc} 
\tablecaption{Reduced $\chi^2$ values for various spectral models
for the quiescence spectra.
\label{tab:softhard}}
\tablehead{ 
\colhead{}    &  \multicolumn{3}{c}{Hard Component} \\
\cline{2-4}
\colhead{Soft Component} & \colhead{mekal} & \colhead{power law + Gaussian} &
\colhead{thermal brems. + Gaussian}} 
\startdata
mekal     \dotfill & 0.78(43) & 0.76 (42) & 0.76(42) \\
power law \dotfill & 0.77(44) & 0.75 (43) & 0.75(43) \\
blackbody \dotfill & 0.80(44) & 0.76 (43) & 0.76(43) \\ 
\enddata 
\tablecomments{Values in parentheses are the degree of freedom.}
\end{deluxetable} 

\clearpage

%%%%%%%%%%%%%%%%%%%%%%%%%%
\begin{deluxetable}{rcccc}
\tablecolumns{5}
\tablewidth{0pc}
\tablecaption{Best-fit parameters of the average quiescent spectra with
various models.\label{tab:fit_qui}}
\tablehead{
\colhead{Parameter} & \colhead{mekal + pl + gau} & 
\colhead{2 mekal I\tablenotemark{a}} &
\colhead{2 mekal II\tablenotemark{b}} &
\colhead{2 mekal III\tablenotemark{c}}
}
\startdata
$N_{\rm H1}$ (10$^{21}$~cm$^{-2}$) \dotfill 
 & $2.8^{+2.8}_{-1.2}$     & $2.8^{+4.2}_{-1.9}$
 & $7.9^{+1.4}_{-7.9}$     & $8.3^{+1.8}_{-2.0}$ \\
$N_{\rm H2}$ (10$^{23}$~cm$^{-2}$) \dotfill
 & $1.9^{+0.4}_{-0.2}$     & $2.0^{+0.8}_{-0.5}$
 & $2.0^{+0.8}_{-0.6}$     & $1.9^{+0.7}_{-0.5}$ \\
photon index \dotfill
 & $2.28^{+0.85}_{-0.81}$  & $\cdots$
 & $\cdots$                & $\cdots$ \\
$kT_1$ (keV) \dotfill
 & $0.53^{+0.11}_{-0.12}$  & $0.54^{+0.40}_{-0.17}$
 & $0.46^{+0.95}_{-0.11}$  & $0.45^{+0.17}_{-0.14}$ \\
$kT_2$ (keV) \dotfill
 & $\cdots$                & $5.59^{+9.11}_{-2.99}$
 & $5.23^{+74.36}_{-2.93}$ & $5.54^{+7.74}_{-2.92}$ \\
$EM_1$ ($10^{56}d_5^2$~cm$^{-3}$) \dotfill
 & 4.64                    & 4.53
 & 4.53                    & 3.91 \\
$EM_2$ ($10^{56}d_5^2$~cm$^{-3}$) \dotfill
 & $\cdots$                & 4.92
 & 5.66                    & 4.98 \\
$Z_1/Z_{\rm C}$\tablenotemark{d} \dotfill
 & $0.01^{+0.03}_{-0.01}$  & $0.01^{+0.11}_{-0.01}$
 & $0.26^{+0.54}_{-0.26}$  & 0.36 (fix) \\
$Z_2/Z_{\rm C}$\tablenotemark{d} \dotfill
 & $\cdots$                & $0.38^{+0.74}_{-0.29}$
 & $0.26^{+0.54}_{-0.26}$  & 0.36 (fix) \\
Line center (keV) \dotfill
 & $6.51\pm 0.23$          & $\cdots$
 & $\cdots$ & $\cdots$ \\
Line norm. (10$^{-6}$phs. cm$^{-2}$s$^{-1}$) \dotfill
 & $8.5^{+4.2}_{-5.0}$     & $\cdots$
 & $\cdots$ & $\cdots$ \\
Equivalent Width (eV) \dotfill
 & $624^{+418}_{-495}$     & $\cdots$
 & $\cdots$ & $\cdots$ \\
$L_{\rm S,\,bol}$\tablenotemark{e} (10$^{33}$ erg s$^{-1}$) \dotfill
 & $\cdots$                & $\cdots$
 & $\cdots$                & $6.0$ \\
$L_{\rm H,\,bol}$\tablenotemark{f} (10$^{33}$ erg s$^{-1}$) \dotfill
 & $\cdots$                & $\cdots$
 & $\cdots$                & $9.4$ \\
$\chi^2/\nu$(dof) \dotfill
 & 0.76(42)                & 0.78(43)
 & 0.86(44)                & 0.84(45)       \\
\enddata
\tablenotetext{a}{Two abundances free independently.}
\tablenotetext{b}{Two abundances free to vary but constrained to be the same.}
\tablenotetext{c}{Both abundances fixed at the outburst value \citep{ued98}.}
\tablenotetext{d}{$Z_{\rm C}$ implies the cosmic abundance \citep{angr89}.}
\tablenotetext{e}{Bolometric luminosity for the soft component at the
distance of 5~kpc.}
\tablenotetext{f}{Bolometric luminosity for the hard component at the
distance of 5~kpc.}
\end{deluxetable}

\clearpage

%%%%%%%%%%%%%%%%%%%%%%%%
\begin{deluxetable}{lccccc}
\tablecolumns{6}
\tablewidth{0pc} 
\tablecaption{Best-fit parameters of the flare spectra with a power law
overlaid with dual edges.
\label{tab:fit_pl_edges}
}
\tablehead{ 
\colhead{Flare Phase} 
& \colhead{flare 1} 
& \colhead{flare 2}
& \colhead{flare 3}
& \colhead{flare 4}
& \colhead{flare 5} 
}
\startdata
$N_{\rm H}$ ($10^{21}{\rm cm}^{-2}$) \dotfill &
4.6 (fixed) & 4.6 (fixed) & 4.6 (fixed) & 4.6 (fixed) & 4.6 (fixed) \\
Photon index &
 $3\pm2$ & $2.0^{+0.2}_{-0.4}$ & $5.3^{+0.5}_{-1.2}$ 
  & $2.34^{+0.13}_{-0.08}$ & $0.5^{+0.2}_{-0.1}$ \\
Norm.\tablenotemark{\dagger} \dotfill &
 $1.4^{+4.3}_{-0.8}$ & $6^{+1}_{-2}$ & $0.47\pm0.02$ 
  & $8.2\pm^{+0.1}_{-0.2}$ & $27.7\pm^{+0.4}_{-0.3}$ \\
$E_{\rm edge1}$ (keV) \dotfill &
 $0.6144$ (fixed) & $0.66\pm0.01$ & $0.65^{+\infty}_{-0.65}$ 
  & $0.67^{+0.01}_{-0.02}$ & $0.773^{+0.002}_{-0.001}$ \\
$\tau_1$\tablenotemark{a} \dotfill &
 $<3$ & $1.1^{+0.1}_{-0.2}$ & $<1.4$ 
  & $1.06^{+0.08}_{-0.05}$ & $3.1^{+0.2}_{-0.1}$ \\
$E_{\rm edge2}$ (keV) \dotfill &
 $0.6\pm0.1$ & $0.76\pm^{+0.02}_{-0.01}$ & $0.67\pm^{+0.01}_{-0.03}$ 
  & $0.767\pm0.002$ & $0.832\pm^{+0.004}_{-0.006}$ \\
$\tau_2$\tablenotemark{b} \dotfill &
 $<78$ & $12^{+6}_{-5}$ & $<2.7$ & $12^{+6}_{-2}$ & $80\pm 10$ \\
$\chi^2/\nu$ (dof) \dotfill &
 0.81 (51) & 0.92 (104) & 0.82 (50) & 0.80 (104) & 1.03 (104) \\
\enddata 
\tablenotetext{\dagger}{Power-law normalization 
in the unit of $10^{-3}$~ph~s$^{-1}$cm$^{-2}$ at 1~keV.}
\tablenotetext{a}{Optical depth of the atomic edge 
at the energy $E_{\rm edge1}$.}
\tablenotetext{b}{Optical depth of the atomic edge 
at the energy $E_{\rm edge2}$.}
\end{deluxetable} 

\clearpage

%%%%%%%%%%%%%%%%%%%%%%%%
\begin{deluxetable}{lcccccc}
\tablecolumns{7}
\tablewidth{0pc} 
\tablecaption{Best-fit parameters of the flare spectra with a blackbody
and edges.
\label{tab:fit_NHfree}
}
\tablehead{ 
\colhead{Flare Phase} 
& \colhead{flare 1} 
& \colhead{flare 2}
& \colhead{flare 3}
& \colhead{flare 4}
& \colhead{flare 5} 
& \colhead{flare 3\tablenotemark{\dagger}}
}
\startdata
$N_{\rm H}$ ($10^{21}\,{\rm cm}^{-2}$) \dotfill
  & $<7.0$ & 2.0$^{+0.1}_{-0.2}$ & $<0.92$ & 1.7$^{+0.1}_{-0.2}$ 
  & 5.7$\pm$0.1 & $6.2^{+2.7}_{-3.9}$ \\
$kT_{\rm bb}$ (keV) \dotfill
  & $0.11^{+0.09}_{-0.03}$ & $0.128^{+0.002}_{-0.003}$
  & $0.062^{+0.002}_{-0.004}$ & $0.132^{+0.003}_{-0.004}$
  & $0.159^{+0.004}_{-0.003}$ & $0.07^{+0.02}_{-0.01}$ \\
$E_{\rm edge\, 1}$ (keV) \dotfill
  & $0.62\pm 0.02$ & $0.758\pm 0.003$ & $\cdots$ & $0.757^{+0.003}_{-0.002}$
  & $0.776^{+0.002}_{-0.001}$ & $0.67\pm 0.02$ \\
$\tau_1$\tablenotemark{a} \dotfill
  & $3.6^{+1.7}_{-1.3}$ & $2.5\pm 0.2$ & $\cdots$ & $2.9\pm 0.2$
  & $3.4\pm 0.1$ & $>2.1$ \\
$E_{\rm edge\, 2}$ (keV) \dotfill
  & $\cdots$ & $0.869^{+0.010}_{-0.009}$ & $\cdots$ & $0.867^{+0.010}_{-0.009}$
  & $0.868^{+0.002}_{-0.003}$ & $\cdots$ \\
$\tau_2$\tablenotemark{b} \dotfill 
  & $\cdots$ & $4.6^{+5.4}_{-2.0}$ & $\cdots$ & $>3.0$
  & $>7.3$ & $\cdots$ \\
$\chi^2_\nu$ (dof) \dotfill
  & 0.98 (51)  & 0.91 (103)  & 0.94 (53)  & 0.86 (103) 
  & 0.89 (129)  & 0.69 (51) \\
\enddata 
\tablenotetext{\dagger}{A single edge is added although no edge model
is acceptable.}
\tablenotetext{a}{Optical depth of the atomic edge 
at the energy $E_{\rm edge1}$.}
\tablenotetext{b}{Optical depth of the atomic edge 
at the energy $E_{\rm edge2}$.}
\end{deluxetable} 

\clearpage

%%%%%%%%%%%%%%%%%%%%%%%%
\begin{deluxetable}{llccccc}
\tablecolumns{7}
\tablewidth{0pc} 
\tablecaption{Best-fit parameters of 
the flare spectra with a blackbody and 
the five edges of \ion{O}{4}--\ion{O}{8}.
\label{tab:fit_NHfix}
}
\tablehead{ 
\colhead{Flare Phase} 
& \colhead{V \& Y\tablenotemark{\dagger}}
& \colhead{flare 1} 
& \colhead{flare 2}
& \colhead{flare 3}
& \colhead{flare 4}
& \colhead{flare 5} 
}
\startdata
$N_{\rm H}$\tablenotemark{\pounds} ($10^{21}$~cm$^{-2}$) \dotfill & $\cdots$ 
 & $<32$ & $4.5\pm0.3$ & $<7.6$
 & $0.8\pm0.3$ & $7^{+2}_{-3}$ \\
$kT_{\rm bb}$ (keV) \dotfill  & $\cdots$ 
 & $0.10^{+0.12}_{-0.02}$ & $0.110\pm 0.002$ & $0.07\pm0.01$
 & $0.115^{+0.003}_{-0.002}$ & $0.113^{+0.007}_{-0.005}$ \\
$E_{\rm edge\,1}$ (keV) \dotfill  & 0.6144
 & $0.60\pm0.03$ & $0.641^{+0.003}_{-0.001}$ & $0.59^{+0.03}_{-0.01}$
 & $0.64^{+0.02}_{-0.05}$ & $0.66^{+0.03}_{-0.01}$ \\
$\tau_1$ \dotfill  & $\cdots$
 & $ < 29$ & $0.2\pm0.1$ & $<1.7$ 
 & $ < 0.1$ & $ < 0.1$ \\
$E_{\rm edge\,2}$\tablenotemark{\P} (keV) \dotfill  & 0.6491
 & $0.64$ & $0.676$ & $0.63$
 & $0.676$ & $0.69$ \\
$\tau_2$ \dotfill  & $\cdots$
 & $ 5^{+24}_{-2}$ & $0.9\pm0.1$ & $< 1.1$
 & $ 0.65\pm0.06$ & $ < 0.2$ \\
$E_{\rm edge\,3}$\tablenotemark{\P} (keV) \dotfill  & 0.6837
 & $0.67$ & $0.713$ & $0.66$
 & $0.712$ & $0.73$ \\
$\tau_3$ \dotfill  & $\cdots$
 & $1$ \tablenotemark{\S} & $ 0.25^{+0.06}_{-0.07}$ & $>2.3$
 & $0.22^{+0.07}_{-0.06}$ & $0.3^{+0.3}_{-0.2}$ \\
$E_{\rm edge\,4}$\tablenotemark{\P} (keV) \dotfill  & 0.7393
 & $0.73$ & $0.771$ & $0.72$
 & $0.770$ & $0.79$ \\
$\tau_4$ \dotfill  & $\cdots$
 & $0.0$ (fixed) & $4.3\pm 0.3$ & $0.0$ (fixed)
 & $4.3\pm0.3$ & $4.9^{+0.3}_{-0.2}$ \\
$E_{\rm edge\,5}$\tablenotemark{\P} (keV) \dotfill & 0.8714
 & $0.86$ & $0.908$ & $0.84$
 & $0.907$ & $0.93$ \\
$\tau_5$ \dotfill  & $\cdots$
 & $0.0$ (fixed) & $ > 84$ & $0.0$ (fixed)
 & $> 2.1$ & $ 109$\tablenotemark{\S} \\
$L_{\rm bol}$\tablenotemark{\ddag} ($10^{38}{\rm erg\;s}^{-1}$) \dotfill 
& $\cdots$
 & $0.3^{+19.7}_{-0.1}$ & $2.2^{+2.7}_{-1.8}$ & $1.3^{+18.7}_{-1.0}$
 & $1.0^{+5.1}_{-0.1}$ & $9.0^{+82.0}_{-8.2}$ \\
$\chi^2_\nu$ (dof) \dotfill & $\cdots$
 & 0.85 (49) & 0.92 (101) & 0.83 (49) 
 & 0.83 (101) & 0.85 (101) \\
\enddata 
\tablenotetext{\pounds}{Hydrogen column density in excess of
$4.6\times 10^{21}{\rm cm}^{-2}$}.
\tablenotetext{\dagger}{Theoretical ionized oxygen edge energies
from \ion{O}{4} through \ion{O}{8} taken from \citet{ver95}.}
\tablenotetext{\ddag}{Distance is assumed to be 5~kpc.}
\tablenotetext{\S}{The depth is unbound within the range $0 - 200$,
and we fixed at the value tabulated.}
\tablenotetext{\P}{Linked to $E_{\rm edge\,1}$ so as to be consistent 
with the theoretical value \citep{ver95} listed in the second column.}
\end{deluxetable} 

\clearpage

%%%%%%%%%%%%%%%%%%%%%%%%
\begin{deluxetable}{ccccccccc}
\tablecolumns{9}
\tablewidth{0pc} 
\tablecaption{Best-fit parameters of the optical light curve fit 
and the resultant distance range.
\label{tab:vsnet}}
\tablehead{ 
%\cline{2-4}
\colhead{$t_{\rm max}$\tablenotemark{\ddag}} 
& \colhead{$m_V^{\rm cosnt}$} 
& \colhead{$m_V^{\rm amp}$}
& \colhead{$\tau_{\rm fold}$}
& \colhead{$t_2$}
& \colhead{$m_V^{\rm max}$} 
& \colhead{$M_{\rm V}^{\rm max}$}
& \colhead{$m_V^{\rm max} - M_V^{\rm max}$\tablenotemark{\dagger}}
& \colhead{Distance} \\
\colhead{(1998 Apr.)} 
& \colhead{} 
& \colhead{}
& \colhead{(d)}
& \colhead{(d)}
& \colhead{} 
& \colhead{}
& \colhead{}
& \colhead{(kpc)}
} 
\startdata
1.0 & 11.80 & 2.98 & 6.97 & 5.80 & 8.83 
	& $-$8.86 & 13.3--15.7 & 4.5--13.7 \\
2.0 & 11.80 & 2.58 & 6.97 & 6.50 & 9.22 
	& $-$8.74 & 13.6--16.0 & 5.2--15.6 \\
3.0 & 11.80 & 2.23 & 6.97 & 7.19 & 9.57 
	& $-$8.64 & 13.8--16.2 & 5.8--17.4
\enddata 
\tablenotetext{\ddag}{Date of the optical maximum.}
\tablenotetext{\dagger}{$A_V$ corrected.}
\end{deluxetable} 

\clearpage

%%%%%%%%%%
% FIGURE 1
%%%%%%%%%%
%
%\clearpage
%

%%%%%%%%%%
% FIGURE 2
%%%%%%%%%%
%
%\clearpage
%

%%%%%%%%%%
% FIGURE 3
%%%%%%%%%%
%
%\clearpage
%

%%%%%%%%%%
% FIGURE 4
%%%%%%%%%%
%
%\clearpage
%

%%%%%%%%%%
% FIGURE 5
%%%%%%%%%%
%
%\clearpage
%

\end{document}